\documentclass[10pt]{article}
\usepackage{amsmath}
\usepackage{amssymb}
\usepackage{graphicx}
\usepackage{multirow}
\usepackage{cite}
\usepackage{color}
\usepackage[labelfont=bf,labelsep=period,justification=raggedright]{caption}

\makeatletter
\renewcommand{\@biblabel}[1]{\quad#1.}
\makeatother

\newcommand{\numcom}{\ensuremath{|\mathcal{C}(\mathcal{G}_t)}|}
\newcommand{\com}{\ensuremath{\mathcal{C}(\mathcal{G}_t)}}

\date{}

\begin{document}

\begin{flushleft}
{\Large
\textbf{Community Structure and the Evolution of Interdisciplinarity in Slovenia's Scientific Collaboration Network}
}\sffamily
\\[3mm]
\textbf{Borut Lu{\v z}ar$^{1}$, Zoran Levnaji{\'c}$^{1}$, Janez Povh$^{1}$, Matja{\v z} Perc$^{2,\ast}$}
\\[2mm]
{\bf 1} Faculty of Information Studies in Novo mesto, Novo mesto, Slovenia\\
{\bf 2} Faculty of Natural Sciences and Mathematics, University of Maribor, Maribor, Slovenia
\\
$^\ast$matjaz.perc@uni-mb.si
\end{flushleft}
\sffamily

\section*{Abstract}
Interaction among the scientific disciplines is of vital importance in modern science. Focusing on the case of Slovenia, we study the dynamics of interdisciplinary sciences from $1960$ to $2010$. Our approach relies on quantifying the interdisciplinarity of research communities detected in the coauthorship network of Slovenian scientists over time. Examining the evolution of the community structure, we find that the frequency of interdisciplinary research is only proportional with the overall growth of the network. Although marginal improvements in favor of interdisciplinarity are inferable during the 70s and 80s, the overall trends during the past 20 years are constant and indicative of stalemate. We conclude that the flow of knowledge between different fields of research in Slovenia is in need of further stimulation.

\section*{Introduction}
Recent research has highlighted the importance of interdisciplinarity for ground breaking discoveries \cite{uzzi2013atypical}. If during the past centuries advances in science were due to disciplinary thinking and the meticulous dissection of different fields of research on the most elementary subdisciplines, it seems now the time may be ripe for the integration of the accumulated knowledge to form a new, and above all a better, understanding of the complex world that has emerged \cite{ball_12}. The push towards interdisciplinary efforts is reflected in the recently released guidelines of the Horizon 2020 -- The EU Framework Programme for Research and Innovation -- and it is also reflected in the agenda of the Slovenian Research Agency, which a decade ago set up a special Expert Body for Interdisciplinary Research to foster the exchange of knowledged and collaboration between disciplines. The question is to what extent these measures are successful in bringing about the desired change, in particular the dissemination and promotion of interdisciplinarity. It is namely not rare that such policies, although being developed with the best intentions, fail. A recently identified example of a similar failure is the development of an integrated European Research Area, which was thought to be a critical component for a more competitive and open European research and development system. But as \cite{chessa2013europe} point out, there has been little integration above global trends in patenting and publication, thus leaving Europe as a collection of national innovation systems rather than an integrated research area.

Here we make use of Slovenia's research history \cite{perc2010growth} and methods for community detection in networks \cite{fortunato2010community} to study the evolution of communities and their interdisciplinarity during the past 50 years. Community detection has gained on popularity as the methodology best suited for analyzing social networks and understanding global human interactions \cite{wassermanXbook94}. The methods for community detection have also been utilized to identify reaction modules in metabolic networks \cite{girvan2002community}, protein structure \cite{szalay2012moduland}, and to study self-organization and identification of web communities \cite{flake2002self}, for example, in addition to the many other aspects of real-life complex systems \cite{newman2006modularity, boccaletti2006complex, palla2007quantifying, fortunato2010community, kenett2012dependency, delpini2013evolution}. Community detection is NP-hard, which gave rise to an array of heuristic methods developed over the past decade \cite{fortunato2010community}. While modularity optimization \cite{newman2004finding} is still employed frequently, the resolution limit \cite{fortunato2007resolution} and the advent of local optimization techniques \cite{lancichinetti2009detecting, LanRadRamFor11} led to massive research efforts being invested into finding, testing, and validating various new methods \cite{palla2005uncovering, lancichinetti2008benchmark, arenas2008analysis, kovacs2010community}. In our paper, we employ three different methods: ``Louvain'' method \cite{BloGuiLamLef08}, the COPRA algorithm \cite{Gre10}, and the OSLOM algorithm \cite{LanRadRamFor11}. As we show in what follows, the study of evolution and interactions among the research communities in Slovenian coauthorship network provides a unique opportunity to observe the coming of age of a country's research system. On the other hand, it allows us to assess the effectiveness of national policies that were installed to promote and foster interdisciplinary research.

Before presenting the main results concerning the community structure and the evolution of interdisciplinarity (see Fig.~\ref{scheme} for the definition of the interdisciplinarity measure), we briefly summarize the key structural properties of Slovenia's scientific collaboration network. There were no more than $30$ scientists with an average of $1.5$ collaborators in the year $1960$, while to date the network consists of over $12609$ individuals that, on average, have $10.9$ collaborators. The network has properties that are typical of ``small worlds'', and its growth is governed by near-liner preferential attachment. In \cite{perc2010growth}, we have shown that there exists a tipping point in time after which the mean distance between authors and the diameter start decreasing, and which coincides with the largest component exceeding $70\%$ of the network size. Time wise, the emergence of the giant connected component and the evolution towards a small world agrees with the introduction of the ``Young Researchers'' program in $1985$, which was backed up by substantiable resources directed towards promoting research in Slovenia. Unfortunately, the introduction of the Expert Body for Interdisciplinary Research to foster the exchange of knowledged and collaboration between disciplines in Slovenia received no such support [instead, modest fractions of resources from other (pure) fields of research were drawn for the establishment], and as we will show in what follows, this has thus far not had the desired impact, neither on the structure of the network nor on interdisciplinary research.

\section*{Results}
\begin{figure}
\centering{\includegraphics[width=15cm]{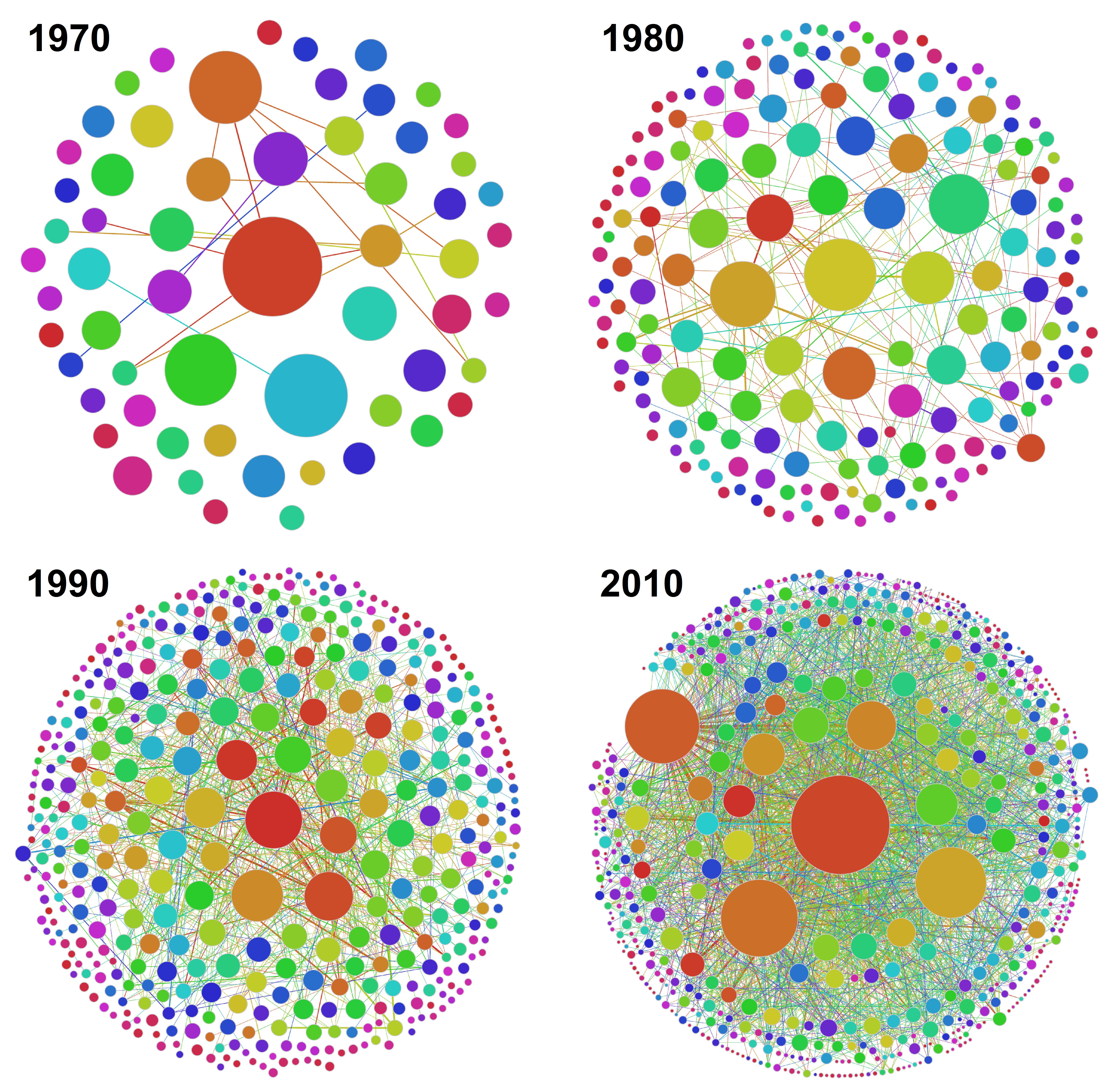}}
\caption{\textbf{Evolution of the community structure of Slovenian scientific coauthorship network, as determined by the COPRA algorithm.} Depicted are the communities and the links between them as obtained for $1970$, $1980$, $1990$ and $2010$. The total number of communities (with five or more members) increases from $60$ ($16$) in the year $1970$ to $933$ ($604$) in the year 2010. The size of the largest community also increases from $23$ to $304$ members during the same time span, and so does the interdisciplinarity from $O(C)=0.484$ to $O(C)=0.739$. Based solely on the analysis of the largest community, one might be tempted to conclude that interdisciplinary research in Slovenia is on the rise. But, as evidenced by the results presented in Fig.~\ref{trends} and Table~\ref{numbers}, this would be a deceitfully optimistic conclusion. The size of each depicted community is proportional to the number of its members, and the thickness of links connecting them is proportional to the logarithm of the number of edges between them. Colors are just to distinguish the communities. Other community detection algorithms yield qualitatively similar results.}
\label{data}
\end{figure}

We begin by showing the evolution of the community structure of Slovenia's coauthorship network in Fig.~\ref{data}, as obtained with the COPRA algorithm. Networks for four representative decades are shown. Results for the 1970 indicate that during the first decade communities were few and practically disconnected from one another. The situation began improving in the 70s and 80s, during which the number of communities as well as the number of links amongst them rose significantly. Since the diameter of the displayed communities is proportional to the number of the members they contain, it can also be observed that the heterogeneity in size also increased significantly during the formative years of the network. This in turn indicates that some communities were more successful in expanding, and that thus some fields grew faster than others, which ultimately gave rise to the strongly heterogeneous Zipf-like distribution of various measures of research productivity and success \cite{perc2010zipf}. The trends of growth and enhanced interrelatedness of communities continue up to the present time, and they are in agreement with the overall growth of the coauthorship network \cite{perc2010growth}. Due to the network size and the related visual limitations, we do not show results for the year 2000 as they are (visually) practically identical to those obtained for the 2010.

A more quantitative view of the growth of the number of communities is attainable with the data presented in Table~\ref{numbers}, where the numbers in brackets denote the number of communities. We show the results obtained with the three considered algorithms, although other methods, including those based on modularity optimization \cite{newman2004finding, newman2004fast}, yield practically identical results. The number of communities, not taking into account those with less than five members, increased by nearly two orders of magnitude during the past $50$ years, with the growth being fairly steady across the examined history. Relatively, the growth was the fastest during the 70s and 80s, but this is likely related to the formation of fundamental research infrastructure and mechanisms of research promotion (e.g. launching the ``Young Researchers'' program in $1985$). During the past two decades, approximately $100$ new communities emerge every five years ($\approx 20$ per year), which fits well with the yearly increase in the network size of about $100-200$ new active researchers (of course not all will go on to give rise to new communities).

\begin{figure}
\centering{\includegraphics[width=12cm]{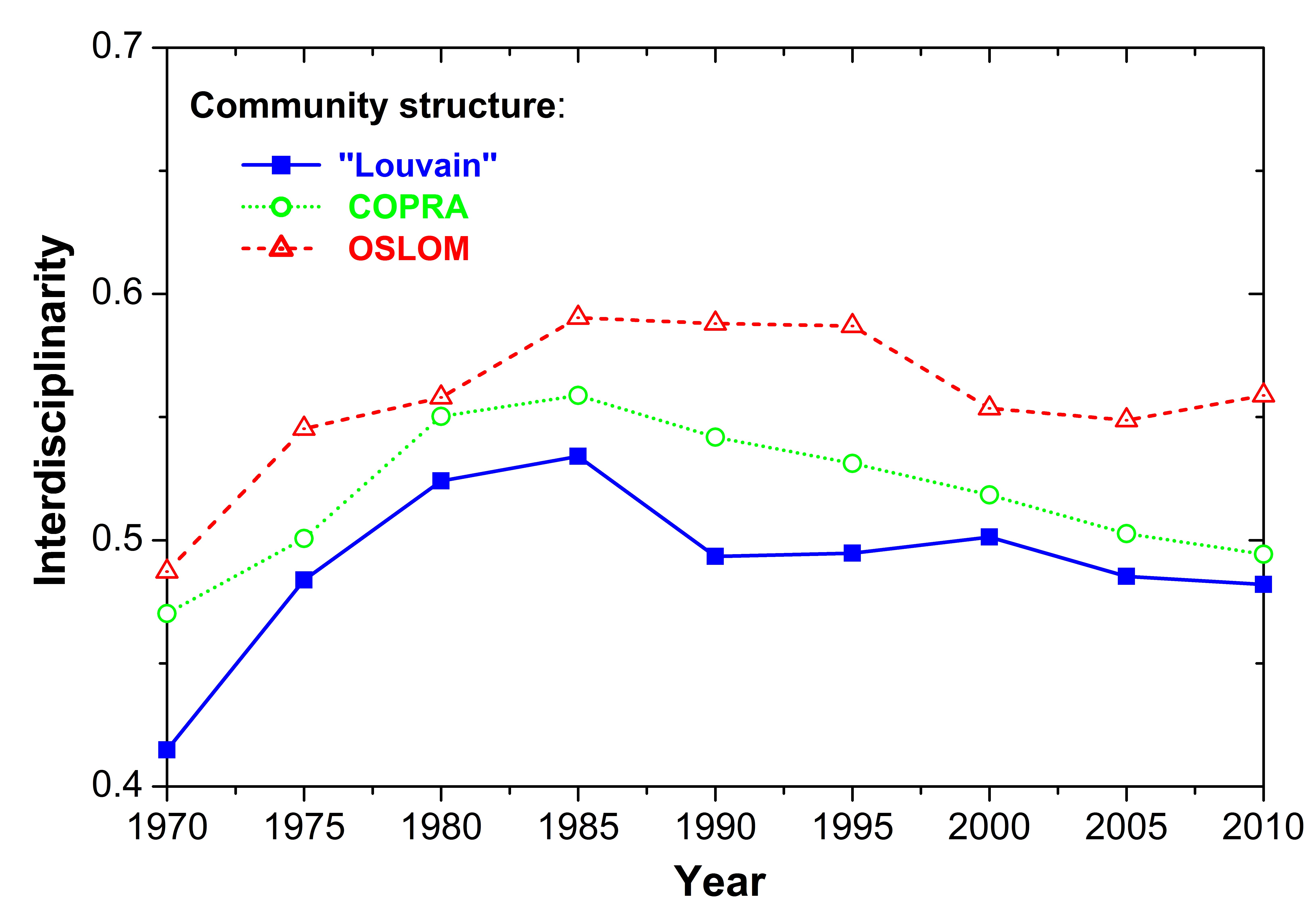}}
\caption{\textbf{Evolution of interdisciplinarity in Slovenian coauthorship network during the past 40 years.} Depicted is the interdisciplinarity measure $O(\mathcal{G}_t)$ defined by Eq.~\ref{internet}, as derived from the communities identified with the ``Louvain Method'', the COPRA algorithm and the OSLOM algorithm. There is a relatively modest increase in interdisciplinary research during the 70s and 80s, but subsequently the upward momentum is lost and the trend even seems to be reverting during the past two decades. The results are largely independent of the methods and measurements used.}
\label{trends}
\end{figure}

In terms of the interdisciplinarity of the research communities, however, the trends are far more bleak. While the number and the size of communities has been increasing, the amount of interdisciplinary research has remained constant. As the numbers in Table~\ref{numbers} show, the average interdisciplinarity of the communities that form Slovenian coauthorship network (see Eq.~\ref{internet}) exhibits slight growth only during the 70s and 80s, while the last two decades have not seen any improvement at all. If anything, the trends seem to be going downward rather than upward. These results are independent of the algorithm for community detection, and they are also independent of the measure of interdisciplinarity. We have tested many different versions of Eqs.~\ref{internet} and \ref{intercomm} without observing appreciable qualitative change. For clarity, we display trends of interdisciplinarity also in Fig.~\ref{trends}, which confirm the stalemate in Slovenia's interdisciplinary research efforts.

Linking the average values of interdisciplinarity of around $0.5$ to the definition of the measure (see also Fig.~\ref{scheme}), we come to the conclusion that the researchers in the majority of communities are from the same field of research, with perhaps one or the other deviation occurring intermittently. A more comprehensive insight into the formation of individual communities is attainable from the distributions of the interdisciplinarity measure of individual communities (see Eq.~\ref{intercomm}), as displayed in Fig.~\ref{dist}. Regardless of the year, there is a peak at $C(O)=0$, which grows proportionally with the total number of communities and the network over the decades. The remaining communities have $0.2 \leq O(C) \leq 0.85$, distributed roughly Gaussian, whereby this part of the distribution grows proportionally in amplitude over the years as well. If we normalize the number of communities for each specific time period, we obtain results depicted in the inset of Fig.~\ref{dist}. It can be observed that all the curves fall onto roughly the same trajectory, the only difference being that during the formative years (the 70s and 80s) there is substantially more noise in the intermediate $O(C)$ region. The latter, however, is mainly due to the small sample size, i.e., the small number of communities on which the statistics is based. These results confirm the conclusions offered by the results presented in Table~\ref{numbers} and Fig.~\ref{trends}, indicating that not much has changed in the interdisciplinarity landscape of Slovenia's research during the past 50 years, despite ample efforts, especially during the last decade, to promote interdisciplinary research. The communities that form spontaneously during the network growth are primarily composed of researchers from a particular field, and only seldom is there a fusion of knowledge from different fields such that each would be representative for the community as a whole. Our analysis also suggests that the links between the communities are predominantly due to institutional relatedness, rather than due to efforts of bridging barriers between the disciplines.

\section*{Discussion}
We have studied the evolution of the community structure and interdisciplinarity in Slovenia's scientific collaboration network during the past 50 years. The SICRIS database offers unique insights into the growth and evolution of a country's research ecosystem, and we find that the one of interdisciplinarity has been in a relative recession during the time span that is subject to our analysis. On the one hand, the fact that interdisciplinary research has been growing proportionally with the overall growth of the collaboration network can be interpreted as a silver lining development. On the other hand, the hope would be that, in the light of the importance of interdisciplinary research and the implemented policies that favor such development, the interdisciplinarity would grow faster than average. Thus, we find that while the network and the number of communities and the links between continue to grow at a steady rate, the amount of interdisciplinary research is stalling or even slightly declining. This invites the conclusion that a healthy and flourishing interdisciplinary research environment in Slovenia is in need of additional and stronger stimulation than it has received thus far. In the future, it would be interesting to conduct similar analysis on larger geographical regions, and to compare how the rate of interdisciplinary research scales with the overall scientific success and productivity. The importance of overlapping communities also merits attention, in particular to test whether the overlap between the different research communities increases over time \cite{overlap}. As pointed out in the Introduction, recent research emphasizes the importance of interdisciplinary efforts for ground breaking discoveries \cite{uzzi2013atypical} as well as for the better management and understanding of our societies \cite{ball_12}, and it thus may well be that the additional support for interdisciplinary research would be quick to pay off, with dividends.

\section*{Methods}
Slovenia has a thoroughly documented research history, made possible by SICRIS -- Slovenia's Current Research Information System -- which hosts complete publication records of all Slovene researchers from the 1960 onwards. We use this database to construct coauthorship networks, where two researchers (considered as network nodes) are connected by an edge if, up to the given year inclusive, they have coauthored at least one paper. The edges are weighted, in the sense that if they coauthored $k$ papers, then the weight of the edge connecting them is $k$. Starting with 1960 and ending with 2010, we construct coauthorship networks by cumulating the edges among the researchers active the time period up to a given year. We term them $\mathcal{G}_t$, where $t$ indicates the ending year. The SICRIS data used are obtained on 14 December 2013.

Starting with no more than $30$ researchers with an average of $1.5$ collaborators in the year $1960$, the network to date consists of 12609 individuals that, on average, have $10.9$ collaborators. The growth of the network is governed by near-linear preferential attachment, giving rise to a log-normal distribution of collaborators per author and small-world properties. For details regarding the network growth and structure, and statistical analysis of the individual scientific indicators, we refer to \cite{perc2010growth, perc2010zipf}.

Next we determine the community structure $\com$ for each network, using three approaches: ``Louvain'' method \cite{BloGuiLamLef08}, the COPRA algorithm \cite{Gre10}, and the OSLOM algorithm \cite{LanRadRamFor11}. We ignore the isolated researchers as well as communities with less than five members. All three algorithms are implemented and freely available on the NetCom Analyzer web page {\color{blue}{www.netcom-analyzer.org}}.

\begin{figure}
\centering{\includegraphics[width=14cm]{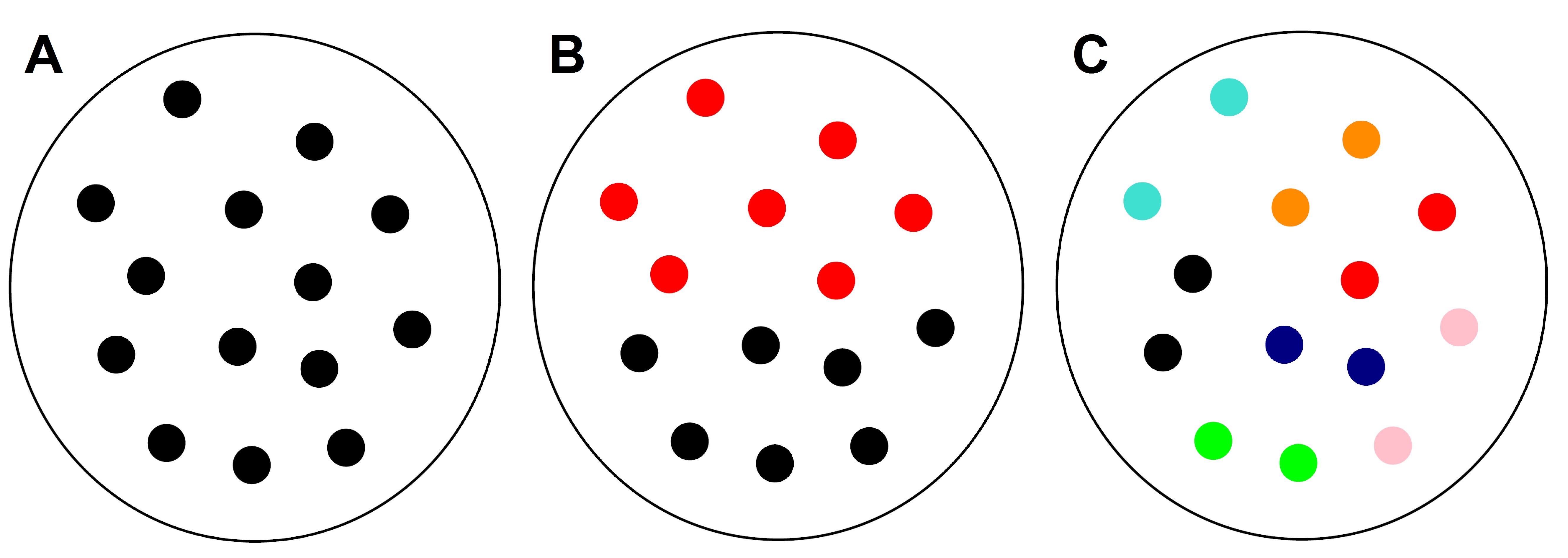}}
\caption{\textbf{Schematic presentation of three communities, illustrating the employed interdisciplinarity measure.} Within the communities, researchers belonging to different
	categories are marked with a different colors. \textbf{A} All researchers work in the research area $x_1$. The seven-component vector is thus
	$\mathcal{I}_{C} = (1,0,0,0,0,0,0)$ and the interdisciplinarity of such a community is, according to Eq.~\ref{intercomm}, $O(C) = 0$. \textbf{B} Researchers
	within this community are evenly spread between areas $x_1$ and $x_4$. The vector is thus $\mathcal{I}_{C} = (\frac{1}{2}, 0, 0,\frac{1}{2},0,0,0)$,
	and the interdisciplinarity of this community is $O(C) = 0.88$. \textbf{C} Here each marked pair of researchers works in a different area,
	thus yielding $\mathcal{I}_{C} = (\frac{1}{7}, \frac{1}{7}, \frac{1}{7}, \frac{1}{7}, \frac{1}{7}, \frac{1}{7}, \frac{1}{7})$ and $O(C) = 1$.
	The edges within communities are not depicted.}
\label{scheme}
\end{figure}

\begin{figure}
\centering{\includegraphics[width=12cm]{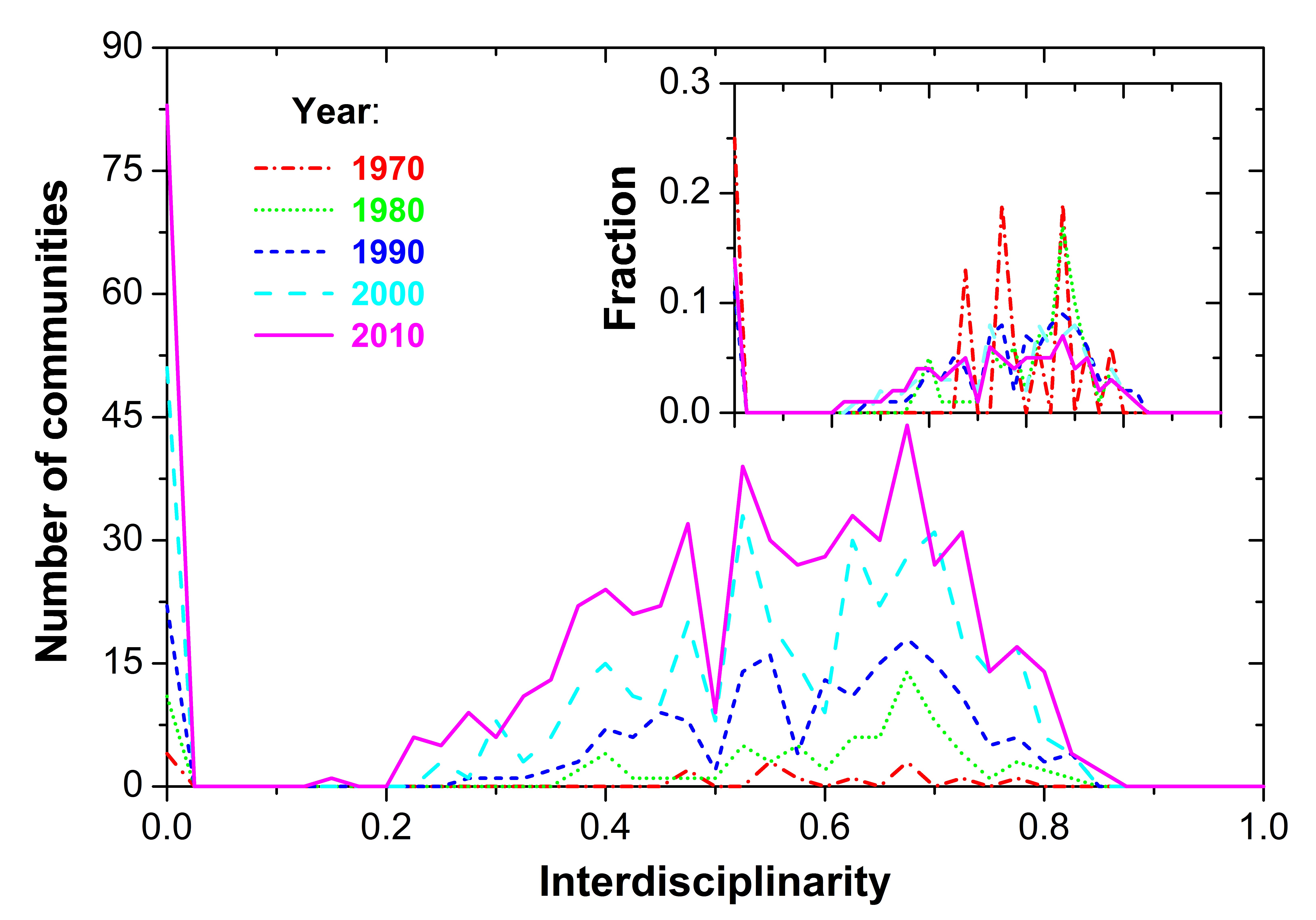}}
\caption{\textbf{Evolution of the distribution of interdisciplinarity within the communities of Slovenia's coauthorship network, as determined by the COPRA algorithm.} The main panel depicts the number of communities with a given interdisciplinarity $O(C)$, while the inset shows the relative fraction of the communities with a given $O(C)$. It can be observed that, regardless of the decade (see figure legend), the majority of communities contain researchers that all work in the same field ($O(C)=0$, see also panel \textbf{A} of Fig.~\ref{scheme}). The sharp peak at $O(C)=0$ is followed by a relatively broad distribution spanning $0.2 \leq O(C) \leq 0.85$, with a maximum (fitted, not shown) at approximately $O(C)=0.6$. This indicates that there are also communities within which researchers work on different field of research, i.e., interdisciplinary communities, but the relative fraction of those changed very little over the years (see inset). Thus, despite ample efforts to promote interdisciplinary research, we arrive at the sobering conclusion that, relatively, the landscape of Slovenia's interdisciplinary research has changed little during the past 50 years. On the up side, we may also conclude that the number of interdisciplinary communities grows proportionally with the total number of communities (which could be interpreted as a positive development), yet the desired global shift towards interdisciplinarity is certainly absent. Other community detection algorithms yield qualitatively similar results.}
\label{dist}
\end{figure}

To each researcher registered in the database, SICRIS associates one or more number(s) between $1$ and $7$, defining her/his primary field(s) of work. These seven top-level categories are: Natural sciences and mathematics, Engineering sciences and technologies, Medical sciences, Biotechnical sciences, Social sciences, Humanities, and Interdisciplinary studies. This seventh category is an attempt of SICRIS to quantify interdisciplinarity, but researchers themselves rarely choose ``Interdisciplinary studies'' as their main field. We therefore designed our own way of measuring interdisciplinarity, rather than simply looking at the number of researchers in this group. We use this classification scheme to quantify the interdisciplinarity of each community $C$. We assign a seven-component vector $\mathcal{I}_C$, where each component represents the fraction of researchers within $C$ belonging to one of the seven categories. The interdisciplinarity of a community $O(C)$ is then defined as
\begin{equation}
O(C) = A \sqrt{1 - \sum_{i=1}^7 x_i^2}\,,
\label{intercomm}
\end{equation}
where $x_i$ is the $i$-th component of $\mathcal{I}_C$ and $A=[1 - (1/7)]^{-0.5}$ is a normalization constant ensuring that $0 \leq O(C) \leq 1$. According to Eq.~\ref{intercomm} $O(C) = 0$ if $x_i=1$ for any of the seven components (in this case all the other components are $0$), and $O(C) = 1$ if every component $x_i$ is equal to $\frac{1}{7}$. To illustrate our quantification scheme, in Fig.~\ref{scheme} we depict three communities, each characterized with a different value of $O(C)$. Lastly, based on the definition of interdisciplinarity for each community $C$, we define the interdisciplinarity of the entire coauthorship network for a given period $\mathcal{G}_t$ as
\begin{equation}
	O(\mathcal{G}_t) = \frac{1}{\numcom} \sum_{C \in \com} O(C)\,.
\label{internet}
\end{equation}
Recall that in $\com$ only the communities with five or more members are present.
\clearpage

\clearpage

\clearpage

\begin{table}	
\centering
\begin{tabular}[c]{|c|c|c|c|c|c|} \hline
	   &  		COPRA				 & ``Louvain''					  &  OSLOM \\
network	   &   $O(\mathcal{G}_t)$~~ ($\numcom$)]	 &  $O(\mathcal{G}_t)$ ~~($\numcom$)		  & $O(\mathcal{G}_t)$ ~~($\numcom$)  \\ \hline		
$\mathcal{G}_{70}$ & 0.470 ~~(16) & 0.415 ~~(12) & 0.487 ~~(14) \\ \hline		
$\mathcal{G}_{75}$ & 0.501 ~~(41) & 0.484 ~~(36) & 0.545 ~~(35) \\ \hline		
$\mathcal{G}_{80}$ & 0.550 ~~(81) & 0.524 ~~(74) & 0.553 ~~(57) \\ \hline		
$\mathcal{G}_{85}$ & 0.559 ~~(118) & 0.534 ~~(117) & 0.590 ~~(76) \\ \hline		
$\mathcal{G}_{90}$ & 0.542 ~~(197) & 0.493 ~~(193) & 0.588 ~~(132) \\ \hline		
$\mathcal{G}_{95}$ & 0.531 ~~(282) & 0.495 ~~(291) & 0.587 ~~(170) \\ \hline		
$\mathcal{G}_{00}$ & 0.518 ~~(395) & 0.501 ~~(429) & 0.554 ~~(222) \\ \hline		
$\mathcal{G}_{05}$ & 0.503 ~~(515) & 0.485 ~~(550) & 0.549 ~~(294) \\ \hline		
$\mathcal{G}_{10}$ & 0.494 ~~(604) & 0.482 ~~(689) & 0.559 ~~(391) \\ \hline			
\end{tabular}
\caption{\textbf{Evolution of interdisciplinarity.} Interdisciplinarity of $O(\mathcal{G}_t)$ with 5 year resolution, and the number of communities with more than five members $\numcom$ (in brackets), during the examined time period, as obtained with the ``Louvain'' method, the COPRA algorithm and the OSLOM algorithm. While the number of communities increases steadily, the average level of interdisciplinarity within them remains fairly constant (see also Fig.~\ref{trends}).}
\label{numbers}
\end{table}

\end{document}